\documentclass[fleqn,11pt,twoside]{article}

\usepackage{amsthm,amsthm,amssymb, color, xcolor,epsfig, graphics, subfigure}

\usepackage{amsmath, graphicx, latexsym, lscape }

\makeatletter
\newcommand{\copyrightnote}[2]{{\renewcommand{\thefootnote}{}
 \footnotetext{\small\it
\begin{flushleft}
 \copyright \ #1   #2  
\end{flushleft}}}}

\newcommand{\Name}[1]{\begin{flushleft}
                       \LARGE \bf #1
                       \end{flushleft}\vspace{-3mm}}

\newcommand{\Author}[1]{\begin{flushleft}
                       \it #1 \end{flushleft}}

\newcommand{\Address}[1]{\begin{flushleft}
                       \it #1 \end{flushleft}}

\newcommand{\Date}[1]{\begin{flushleft}
                      \small  \it #1 \end{flushleft}}

\newcommand{\evenhead}{Author \ name}
\newcommand{\oddhead}{Article \ name}

\renewcommand{\@evenhead}{
\hspace*{-3pt}\raisebox{-15pt}[\headheight][0pt]{\vbox{\hbox to \textwidth
{\thepage \hfil \evenhead}\vskip4pt \hrule}}}
\renewcommand{\@oddhead}{
\hspace*{-3pt}\raisebox{-15pt}[\headheight][0pt]{\vbox{\hbox to \textwidth
{\oddhead \hfil \thepage}\vskip4pt\hrule}}}
\renewcommand{\@evenfoot}{}
\renewcommand{\@oddfoot}{}

\long\def\@makecaption#1#2{%
  \vskip\abovecaptionskip
  \sbox\@tempboxa{\small \textbf{#1.}\ \ #2}%
  \ifdim \wd\@tempboxa >\hsize
    {\small \textbf{#1.}\ \ #2}\par
  \else
    \global \@minipagefalse
    \hb@xt@\hsize{\hfil\box\@tempboxa\hfil}%
  \fi
  \vskip\belowcaptionskip}

\newcommand{\JNMPnumberwithin}[3][\arabic]{%
  \@ifundefined{c@#2}{\@nocounterr{#2}}{%
    \@ifundefined{c@#3}{\@nocnterr{#3}}{%
      \@addtoreset{#2}{#3}%
      \@xp\xdef\csname the#2\endcsname{%
        \@xp\@nx\csname the#3\endcsname .\@nx#1{#2}}}}%
}

\newcommand{\resetfootnoterule} {
  \renewcommand\footnoterule{%
  \kern-3\p@
  \hrule\@width.4\columnwidth
  \kern2.6\p@}
}

\renewcommand{\footnoterule}{}

\makeatother

\theoremstyle{definition}

\begin{document}

\renewcommand{\evenhead}{ {\LARGE\textcolor{blue!10!black!40!green}{{\sf \ \ \ ]ocnmp[}}}\strut\hfill 
B Konopelchenko, C Rogers and P Amster
}
\renewcommand{\oddhead}{ {\LARGE\textcolor{blue!10!black!40!green}{{\sf ]ocnmp[}}}\ \ \ \ \   
On an integrable 2+1-dimensional extended Dym equation
}

\thispagestyle{empty}
\newcommand{\FistPageHead}[3]{
\begin{flushleft}
\raisebox{8mm}[0pt][0pt]
{\footnotesize \sf
\parbox{150mm}{{Open Communications in Nonlinear Mathematical Physics}\ \  \ {\LARGE\textcolor{blue!10!black!40!green}{]ocnmp[}}
\ \ Vol.6 (2026) pp
#2\hfill {\sc #3}}}\vspace{-13mm}
\end{flushleft}}

\FistPageHead{1}{\pageref{firstpage}--\pageref{lastpage}}{ \ \ Article}

\strut\hfill

\strut\hfill

\copyrightnote{The author(s). Distributed under a Creative Commons Attribution 4.0 International License}

\Name{On an integrable 2+1-dimensional extended  Dym equation: Lax pair, $\bar{\partial}$-dressing scheme and modulation}

\Author{Boris Konopelchenko$^1$, Colin Rogers$^2$ and Pablo Amster$^3$}

\Address{{$^1$ INFN, Sezione di Lecce, Italy. }
{{\small Email:    {konopelchenkob@gmail.com}}}
\smallskip 

{$^2$ School of Mathematics and Statistics,}
{University of New South Wales, Sydney, Australia.}
{{\small Email: {c.rogers@unsw.edu.au}}}

\smallskip

{$^3$ Depto. de Matem\'atica, Facultad de Ciencias Exactas y Naturales,} {Universidad de Buenos Aires \& IMAS-CONICET.}
   {Ciudad Universitaria. Pabell\'on I (1428), Buenos Aires, Argentina}
{
   {\small Email: pamster@dm.uba.ar}
}}

\Date{Received January 15, 2026; Accepted January 26, 2026}

\setcounter{equation}{0}

\smallskip

\noindent
{\bf Citation format for this Article:}\newline
 Boris Konopelchenko, Colin Rogers and Pablo Amster, 
On an integrable 2+1-dimensional extended Dym equation: Lax pair, $\bar{\partial}$-dressing scheme and modulation,
{\it Open Commun. Nonlinear Math. Phys.}, {\bf 6}, ocnmp:17315, \pageref{firstpage}--\pageref{lastpage}, 2026.

\strut\hfill

\noindent
{\bf The permanent Digital Object Identifier (DOI) for this Article:}\newline
{\it 10.46298/ocnmp.17315}

\strut\hfill

\begin{abstract}
\noindent 
In 1+1-dimensions, an extension of the canonical solitonic Dym equation has previously been derived both in a geometric torsion evolution context and in the analysis of peakon solitonic phenomena in hydrodynamics. Here, a novel 2+1-dimensional S-integrable extended Dym-type equation is introduced. A Lax pair is constructed and an associated $\bar{\partial}$-dressing scheme detailed. Integrable modulated versions of the 2+1-dimensional extended Dym equation are generated via application of a class of involutory transformations with genesis in classical Ermakov theory.
\end{abstract}

\label{firstpage}


\section{Introduction}

The extended 1+1-dimensional Dym equation, namely
\begin{equation} \label{1}
u_t+2\partial_x(1-\partial_{xx})\left(\dfrac{1}{u^{1/2}}\right)=0 \end{equation}
has its origin in the seminal study of peaked solitonic phenomena in hydrodynamics by Camassa and Holm \cite{rcdh83}. In \cite{wscr99}, the nonlinear evolution equation
\begin{equation} \label{2}
\partial \tau/\partial b=\partial\left[\dfrac{1}{\kappa}(\tau^{-1/2})_{ss}-\tau^{3/2}+\kappa \tau^{-1/2}\right]/\partial s \end{equation}
was derived via an intrinsic geometric representation in a description of the evolution of the torsion $\tau(s,b)$ of an inextensible curve of constant curvature $\kappa$ which executes purely binormal motion. Therein $\partial/\partial b$ and $\partial/\partial s$ indicate derivatives in binormal and arcwise directions respectively. It is remarked that such a geometric formalism had been previously applied in magnetohydrodynamics to uncover certain underlying geodesic structure in \cite{crjk74}.

The canonical Dym equation
\begin{equation} \label{3}
\partial \tau/\partial b+2[\tau^{-1/2}]_{sss}=0 \end{equation}
and extended Dym equation \eqref{1} may both, in turn, be retrieved via \eqref{2} on application of appropriate scaling and limiting processes.

It is noted that, the reciprocal transformation
\begin{equation} \label{4}
\begin{array}{c} dx^*=udx-2(1-\partial_{xx})(u^{-1/2})dt, \quad t^*=t, \\[2mm]
u^*=u^{-1}  \end{array} \end{equation}
applied to the extended Dym equation \eqref{1} results in the conservation law
\begin{equation} \label{5}
u^*_{t^*}-2\partial_{x^*}[(u^{*-1/2})_{x^*x^*}+u^{*-3/2}]=0. \end{equation}
The latter, like the Dym and extended Dym equations, may be obtained via appropriate reduction as a specialisation of the torsion evolution equation \eqref{2}. Moreover, with $u^*=v_{x^*}$ and a temporal scaling, \eqref{5} yields a basic conservation law for the Cavalcante-Tenenblat equation \cite{jckt88}
\begin{equation} \label{6}
v^*_{t^*}=(v_{x^*}^{-1/2})_{x^*x^*}+v_{x^*}^{3/2} \end{equation}
which, in turn, may be set in the context of the pseudo-spherical surface theory of Chern and Tenenblat \cite{sckt86}.

That the general torsion evolution equation \eqref{2} is S-integrable in the sense of Calogero \cite{fc93} was established by means of a novel reciprocal link in \cite{wscr99} to the solitonic m$^2$KdV equation as originally introduced by Fokas \cite{af80}.

Nonlinear moving boundary problems of Stefan-type for the extended Dym equation \eqref{1} have been shown in \cite{cr17} to be amenable to exact solution via Painlev\'e II symmetry reduction and iterated application of a B\"acklund transformation. The procedure adopted involved linkage of this class of moving boundary problems to an associated exactly solvable class for the canonical solitonic Dym equation. Moving boundary problems for the latter and its reciprocal associates has previously been analysed in \cite{cr15}. Reciprocal transformations linking the canonical AKNS and WK1 inverse scattering schemes have been set down in \cite{crpw84}. Therein, in addition, novel invariance of the Dym equation under a reciprocal transformation was established. This result was extended to the integrable Dym hierarchy in \cite{crmn86}. The reciprocal invariance admitted by the Dym equation, in principle, may be applied to generate additional associated classes of solvable nonlinear boundary problems for the extended Dym equation.

A 2+1-dimensional integrable version of the canonical Dym equation was originally introduced by Konopelchenko and Dubrovsky in \cite{bkvd84}. It was subsequently linked in \cite{cr87} via a 2+1-dimensional reciprocal-type equation \cite{cr86} to a singularity manifold equation which results from a Painlev\'e integrability test to the Kadomtsev-Petviashvili equation of shallow water hydrodynamics (\cite{bkvp70}).

In 1+1-dimensions, the canonical integrable Dym equation and extended Dym equation \eqref{1} were shown in \cite{hdmp98} to be linked via a change of dependent and independent variables. Here, an analogous procedure applied to the 2+1-dimensional Dym equation of \cite{bkvd84} leads to a novel integrable 2+1-dimensional extension of the latter, namely
\begin{equation} \label{7}
u_t+2\partial_x(1-\partial_{xx})\left(\dfrac{1}{u^{1/2}}\right)+6u^2[u^{-1}\partial^{-1}_x(u^{1/2})_y]_y=0. \end{equation}
It is established here that an appropriate avatar of the latter is amenable to a variant of the standard $\partial$-bar dressing method as originally introduced by Zakharov and Shabat \cite{vzas74} and with extensions detailed in \cite{bk92}.

\section{On Variants of the 2+1-Dimensional Dym Equation: 
Associated Linear Representations}

The 2+1-dimensional Dym equation as introduced in \cite{bkvd84}, namely
\begin{equation} \label{8}
r_t=r^3r_{zzz}+\frac{3}{r}\left(r^2\partial^{-1}_z\left(\frac{r_y}{r^2}\right)\right)_y=0 \end{equation},
admits a corresponding Lax pair
\begin{equation}\begin{array}{c} \label{9}
\Psi_y+r^2\Psi_{zz}=0, \\[2mm]
\Psi_t-4r^3\Psi_{zzz}-6r^2\left[r_z-\partial^{-1}_z\left(\dfrac{r_y}{r^2}\right)\right]\Psi_{zz}=0. \end{array} \end{equation}
Under the change of variables $z\rightarrow x,\ r\rightarrow V$ with
\begin{equation} \label{10}
z=e^{-x}, \end{equation}
\begin{equation} \label{11}
r=e^{-x}V(x,y,t) \end{equation}
the 2+1-dimensional Dym equation \eqref{8} and linear representation \eqref{9} become, in turn
\begin{equation} \label{12}
V_t=V^3(V_x-V_{xxx})+\frac{3}{V}\left[V^2\partial^{-1}_x\left(\frac{1}{V}\right)_y\right]_y \end{equation}
and
\begin{equation}\begin{array}{c} \label{13}
\phi_t+V^2(\phi_{zz}+\phi_x)=0, \\[2mm]
\phi_t+4V^3(\phi_{xxx}+3\phi_{xx}+2\phi_x)-6V^2\left[V-V_x-\partial^{-1}_x\left(\dfrac{1}{V}\right)_y\right](\phi_{xx}+\phi_x)=0 \end{array} \end{equation}
wherein $\phi(x,y,t)=\Psi(t,z,y)$.
In terms of the variable
\begin{equation} \label{14}
u=\frac{1}{V^2}, \end{equation}
\eqref{12} constitutes the 2+1-dimensional extended Dym equation \eqref{7} which generalises explicitly the 1+1-dimensional Camassa-Holm equation \eqref{1}.

It is now convenient to rewrite \eqref{12} as the system
\begin{subequations}
 \begin{gather} 
 V_t=V^3(V_x-V_{xxx})+\dfrac{3}{V}(V^2\rho)_y \label{15}\\
V_y=-V^2\rho_x
\end{gather}
\end{subequations}
and the linear representation \eqref{13} as
\begin{equation}\begin{array}{c} \label{16}
\phi_y+V^2(\phi_{xx}+\phi_x)=0, \\[2mm]
\phi_t+4V^3\phi_{xxx}+6V^3\phi_{xx}+2V^3\phi_x+6V^2(V_x+\rho)(\phi_{xx}+\phi_x)=0. \end{array} \end{equation}
The conservation form
\begin{equation} \label{17}
\rho_x-\left(\frac{1}{V}\right)_y=0 \end{equation}
of \eqref{15} allows the introduction of the potential $W$ such that
\begin{equation} \label{18}
W_x=\frac{1}{V}, \quad W_y=\rho. \end{equation}
In terms of $W$, the 2+1-dimensional Dym equation admits the potential representation
\begin{equation} \label{19}
W_{xt}=-\frac{W_{xxxx}}{W^3_x}+6\frac{W_{xx}W_{xxx}}{W^2_x}+\frac{W_{xx}}{W^3_x}-6\frac{W^3_{xx}}{W^3_x}+6W_yW_{xy}-3W_xW_{yy}. \end{equation}
It is remarked that $W$ is analogous to potentials previously introduced in 2+1-dimensional soliton theory, notably for the canonical Kadomtsev-Petviashvili equation. The potential equation in $W$ is seen to be invariant under the scaling $x\rightarrow x'=\lambda x, \lambda \in \mathbb{R}$.

\section{Linear Representation Analysis}

The linear representation \eqref{16} is now considered with the ansatz
\begin{equation} \label{20}
\phi(x,y,t;\lambda)=\mu(x,y,t;\lambda)e^{F(x,y,t;\lambda)} \end{equation}
wherein $e^F$ is the solution of the system \eqref{16} in the limit $x^2+y^2\rightarrow\infty$ while $\mu$ is normalized according to the condition $\mu(x,y,t;\lambda)\rightarrow1$ as $\lambda\rightarrow\infty$. In addition, in order to proceed, conditions
\begin{equation} \label{21}
V(x,y,t)\rightarrow V_0, \quad \rho(x,y,t)\rightarrow\rho_0 \end{equation}
\begin{equation} \label{22}
V_0,\rho_0 \in \mathbb{R} \end{equation}
are imposed as $x^2+y^2\rightarrow\infty$. In terms of the potential $W$ this requires that
\begin{equation} \label{23}
W(x,y,t)-\left(\frac{1}{V_0}x+\rho_0y\right) \rightarrow 0 \quad \text{as} \quad x^2+y^2\rightarrow\infty. \end{equation}

It is noted that $\dfrac{1}{V_0}x+\rho_0y$ is an admissible time-independent solution of \eqref{19}.

As $x^2+y^2\rightarrow\infty$, the system \eqref{16} becomes
\begin{equation}\begin{array}{c} \label{24}
\phi_{0y}+V^2_0(\phi_{0xx}+\phi_{0x})=0, \\[2mm]
\phi_{0t}+4V^3_0\phi_{0xxx}+6V^3_0+2V^3_0\phi_{0x}+6V^2_0\phi_0(\phi_{0xx}+\phi_{0x})=0 \end{array} \end{equation}
where $\phi\rightarrow\phi_0$ as $x^2+y^2\rightarrow\infty$.
In the $\phi(x,y,t;\lambda)$ representation \eqref{20}, the relation
\begin{equation} \label{25}
F=i\frac{1}{\lambda}\tilde{x}+\frac{1}{\lambda^2}\tilde{y}+4i\left(\frac{1}{\lambda^3}\right)\tilde{t} \end{equation}
is introduced wherein
\begin{equation}\begin{array}{c} \label{26}
\tilde{x}=x-V^2_0y-2V^2_0(V_0+3\rho_0)t, \quad \tilde{y}=V^2_0y+6V^2_0(V_0+\rho_0)t, \\[2mm]
\tilde{t}=V^3_0t. \end{array} \end{equation}
In the sequel, with specialisations $V_0=1,\rho_0=-1$ one has
\begin{equation} \label{27}
F=\frac{i}{\lambda}x+\left(\frac{1}{\lambda^2}-\frac{i}{\lambda}\right)y+\left(\frac{4i}{\lambda^3}+\frac{4i}{\lambda}\right)t \end{equation}
and
\begin{equation} \label{28}
\tilde{x}=x-y+4t, \quad \tilde{y}=y, \quad \tilde{t}=t. \end{equation}
The $\phi$ representation requires that
\begin{equation} \label{29}
\frac{1}{\lambda^2}(1-V^2)\mu+\frac{1}{\lambda}[2iV^2\mu_x-i(1-V^2)\mu]+\mu_y+V^2(\mu_{xx}+\mu_x)=0. \end{equation}
It is now assumed that, in {a} neighbourhood of $\lambda=0,\mu(x,y,t;\lambda)$ adopts the form
\begin{equation} \label{30}
\mu(x,y,t;\lambda)=\sum_{n=-N}\lambda^n\mu_n(x,y,t) \end{equation}
for some positive integer $N$. On substitution of the Laurent series \eqref{30} into \eqref{29} it is readily shown that $\mu$ necessarily has an essential singularity at $\lambda=0$. This impediment can be overcome by introducing the ansatz
\begin{equation} \label{31}
F=\frac{1}{\lambda}f(x,y,t)+\frac{1}{\lambda^2}y+\frac{4i}{\lambda^3}t, \end{equation}
where $f$ satisfies the necessary smoothness requirements. 

\section{The $\bar{\partial}$-Dressing Procedure}

The $\bar{\partial}$-dressing method had its genesis in the nonlocal $\bar{\partial}$-problem
\begin{equation} \label{32}
\frac{\partial\chi}{\partial\bar{\lambda}}=\displaystyle{\iint\limits_{C}} d\lambda' \mathrm{\Lambda} d\bar{\lambda}'  \chi(\lambda', \bar{\lambda}' )R(\lambda' , \lambda) \end{equation}
wherein $\chi$ and $R$ are scalar functions. It is assumed that equation \eqref{32} is uniquely solvable and that $\chi$ is normalised canonically, that is, $\chi\rightarrow 1$ as $\lambda\rightarrow\infty$. A range of S-integrable equations correspond to various dependences of the $\bar{\partial}$-data on the independent variables.

It is assumed here that
\begin{equation}\begin{array}{l} \label{33}
R(\lambda', \bar{\lambda}';\lambda, \bar{\lambda};x,y,t) \\[2mm]
\qquad\qquad = R_0(\lambda', \bar{\lambda}';\lambda, \bar{\lambda})e^{F(\lambda', x,y,t)-F(\lambda, x,y,t)} \end{array} \end{equation}
where $F$ is given by \eqref{31}.

The operators $D_x, D_y, D_t$ are now introduced according to
\begin{equation}\begin{array}{c} \label{34}
D_x=\partial_x+\dfrac{i}{\lambda}f_x, \quad D_y=\partial_y+\dfrac{1}{\lambda^2}+\dfrac{i}{\lambda}f_y, \\[2mm]
D_t=\partial_t+4\dfrac{i}{\lambda^3}+\dfrac{i}{\lambda}f_t. \end{array} \end{equation}
The principal step in the $\bar{\partial}$-dressing method is the construction of operators of the form
\begin{equation} \label{35}
L_1=\sum_{L,m,n}U_{L,m,n}D^n_xD^m_yD^L_t \end{equation}
which obey the condition
\begin{equation} \label{36}
\left[\frac{\partial}{\partial x}, L_1 \right] \chi=0 \end{equation}
together with
\begin{equation} \label{37}
L_1\chi(\lambda)\rightarrow0 \quad \text {as} \quad \lambda\rightarrow\infty. \end{equation}
Such operators yield linear equations
\begin{equation} \label{38}
L_i\chi=0 \end{equation}
the compatibility conditions for which are equivalent to nonlinear integrable PDEs for the coefficients therein.

It is remarked that the operators $D_x, D_y$ and $D_t$ in \eqref{34} adopt the same form as those that are used in connection with the standard 2+1-dimensional Dym equation. However, in the present extended 2+1-dimensional Dym context another class of $f$ therein is required. In terms of $D_x, D_y$ and $D_t$ one obtains the pair 
\begin{equation}\begin{array}{c} \label{39}
D_y\chi+V^2(D^2_x\chi+D_x\chi)=0, \\[2mm]
D_t\chi+4V^3D^3_x\chi+6V^3D^2_x\chi+2V^3D_x\chi+6V^2(V_X+\rho)(D^2_x\chi+D_x\chi)=0 \end{array}\end{equation}
which, with $\phi=\chi e^F$, coincides with that in \eqref{16}. Here, these equations admit solutions regular at $\lambda=0$, that is, with
\begin{equation}\begin{array}{c} \label{40}
\chi(\lambda)=\chi_0(x,y,t)+\lambda\chi_1(x,y,t)+\lambda^2\chi_2(x,y,t)+ ... \\[2mm]
\text{as} \quad \lambda\rightarrow 0 \end{array} \end{equation}
These, in turn, may be shown to provide relations between $f$ and the variables $V, \rho$. Thus, on use of the expressions \eqref{34} the linear system \eqref{39}, \textit{in extenso} yields
\begin{equation}\begin{array}{c} \label{41}
\chi_y+\dfrac{1}{\lambda^2}\chi+\dfrac{i}{\lambda}f_y\chi+V^2\left[\chi_{xx}+\dfrac{i}{\lambda}f_{xx}\chi+\dfrac{2i}{\lambda}f_x\chi_x \right. \\[4mm]
\left. -\dfrac{1}{\lambda^2}f^2_x\chi+\chi_x+\dfrac{i}{\lambda}f_x\chi\right]=0, \end{array} \end{equation}
\begin{equation}\begin{array}{l} \label{42}
\chi_t+\dfrac{4i}{\lambda^3}\chi+\dfrac{i}{\lambda}f_t\chi+4V^3\left[\chi_{xxx}+\dfrac{i}{\lambda}f_{xxx}\chi\right. \\[4mm]
\qquad\qquad \left. +\dfrac{3i}{\lambda}f_{xx}\chi_x+\dfrac{3i}{\lambda}f_x\chi_{xx}-\dfrac{3}{\lambda^2}f_xf_{xx}\chi-\dfrac{3}{\lambda^2}f^2_x\chi_x-\dfrac{i}{\lambda^3}f^3_x\chi\right] \\[4mm]
\qquad\qquad+6V^3\left[\chi_{xx}+\dfrac{i}{\lambda}f_{xx}\chi+\dfrac{2i}{\lambda}f_x\chi_x-\dfrac{1}{\lambda^3}f^2_x\chi\right]+2V^3(\chi_x+\dfrac{i}{\lambda}f_x\chi) \\[4mm]
\qquad\qquad +6V^2\rho\left[\chi_{xx}+\dfrac{i}{\lambda}f_{xx}\chi+\dfrac{2i}{\lambda}f_x\chi_x-\dfrac{1}{\lambda^2}f^2_x\chi+\chi_x+\dfrac{i}{\lambda}f_x\chi\right]=0. \end{array} \end{equation}
Absence of poles in the preceding pair \eqref{41}, \eqref{42} leads to the relations
\begin{equation} \label{43}
V=\frac{1}{f_x}, \end{equation}
\begin{equation} \label{44}
\rho=f_y \end{equation}
together with
\begin{equation} \label{45}
f_y+\frac{1}{f_x}+\frac{f_{xx}}{f^2_x}+\frac{2\chi_{0x}}{f_x\chi_0}=0, \end{equation}
\begin{equation} \label{46}
f_t-\frac{9}{4}f^2_y+\frac{1}{2f^2_x}+\frac{f_{xxx}}{f^3_x}-\frac{3}{2}\frac{f^2_{xx}}{f^4_x}+3\partial^{-1}_x(f_xf_y)_y=0. \end{equation}
It is noted that the latter two equations
contain additional terms $\dfrac{1}{f_x}$ in \eqref{45} and $\dfrac{1}{2f^2_x}$ in \eqref{46}, respectively, to the corresponding system obtained for the standard 2+1-dimensional Dym equation of \cite{bkvd84}. Here, the pair \eqref{45}, \eqref{46} completely characterizes the   function $f(x,y,t)$ associated with the $\bar{\partial}$-dressing procedure for the 2+1-dimensional extended Dym equation \eqref{12}.

It is of interest to note that \eqref{45}, \eqref{46} together imply that
\begin{equation} \begin{array}{c} \label{47}
f_{tx}-6f_yf_{xy}+3f_xf_{yy}+f_{xxxx}/f^3_x-6f_{xx}f_{xxx}/f^4_x \\[4mm]
-f_{xx}/f^3_x+6f^3_{xx}/f^3_x=0 \end{array} \end{equation}
which, with $f=W$, is nothing but the 2+1-dimensional extended Dym potential equation \eqref{19}. Indeed, \eqref{47} may be regarded as a novel alternative avatar of the 2+1-dimensional extended Dym equation as derived in connection with its associated $\bar{\partial}$-dressing procedure.

\section{Modulation}


\subsection{A Class of 2+1-Dimensional Involutory Transformations}

In \cite{wocr93}, a conjugation of reciprocal and gauge transformations was applied to link the 2+1-dimensional Dym, modified Kadomtsev-Petviashvili and Kadomtsev-Petviashvili canonical triad of solitonic hierarchies and their S-integrability properties. Invariance of the canonical base 2+1-dimensional Dym equation was shown in \cite{afcr01} to encode a linear representation for an associated eigenfunction. This result was applied therein to analyse initial-boundary value problems for classes of 2+1-dimensional evolution equations via a procedure originally introduced in \cite{af97}.

In modern soliton theory, reciprocal-type B\"acklund transformations associated with admitted conservation laws were originally introduced in \cite{jkcr82}. Conjugation was made therein with the classical Bianchi theorem of pseudo-spherical surface theory whereby multi-soliton solutions may be generated iteratively in an algorithmic manner. Reciprocal transformations have subsequently proved to have extensive applications in soliton theory [qv \cite{crws02}, \cite{cr22} and literature cited therein]. In nonlinear continuum mechanics, reduction to analytically tractable canonical systems via reciprocal transformations has been detailed in \cite{crws82,crwa89}. In relativistic gasdynamics novel invariance under multi-parameter reciprocal transformations of canonical systems has recently been derived in \cite{crtr20,crtrws20,crtr24}.

Here, novel modulated versions of both the 2+1-dimensional Dym and the Kadomtsev-Petviashvili equations are derived which are reducible to their unmodulated canonical S-integrable counterparts via a class of involutory-type transformations. The type of involutory transformation applied has its genesis in an autonomisation procedure for the coupled Ermakov-Ray-Reid system as detailed in \cite{cacrurao90}. Ermakov-type systems have diverse physical applications in such areas as, \textit{inter alia}, nonlinear optics, oceanographic eddy theory and pulsrodic phenomena in magneto-gasdynamics, (q.v. \cite{crws18} and literature cited therein). \\

\subsection{Modulation via Involutory Transformations}

 \subsubsection{Spatial Modulation} 

The S-integrable 2+1-dimensional Dym equation \cite{bkvd84}
\begin{equation} \label{48}
(r^{-2})_t+2r_{xxx}+6r^{-4}[\ r^2\partial^{-1}_x(r_y/r^2)\ ]_y=0 \end{equation}
on introduction of the transformation with
\begin{equation} \begin{array}{c} \tag{$\mathcal R^*$}\label{49}
\begin{array}{c} dx^*=\rho^{-2}(x)dx\ , \qquad r^*=\rho^{-1}(x)r, \\[2mm]
y^*=y\ , \quad t^*=t\ , \qquad\qquad\qquad\qquad\qquad 
\rho^*=\rho^{-1} \end{array} \end{array} \end{equation}
produces a class of associated spatially modulated 2+1-dimensional Dym equations, namely
\begin{equation} \begin{array}{c} \label{50}
(r^{*-2})_{t^*}+2\rho^{*-2}(\rho^{*2}\partial/\partial x^*)(\rho^{*2}\partial/\partial x^*)(\rho^{*2}\partial/\partial x^*)(r^*/\rho^*) \\[2mm]
+ 6r^{*-4}\left[r^{*2}\partial^{-1}_{x*}\left(\dfrac{1}{\rho^*}r^*_{y^*}/r^{*2}\right)\right]_{y^*}=0\ . \end{array} \end{equation}
It is seen that, under $\mathcal R^*$
%
\begin{equation} \begin{array}{c} \label{51}
dx^{**}=\rho^{*-2}dx^*=dx\ , \quad r^{**}=\rho^{*-1}r^*=r \\[2mm]
y^{**}=y\ , \quad t^{**}=t, \\[2mm]
\rho^{**}=\rho \end{array} \end{equation}
whence, the key involutory property $(\mathcal{R^*})^{2}=\mathrm{I}$ holds.

The canonical solitonic Kadomtsev-Petviashvili equation, namely \cite{bkvp70}
\begin{equation} \label{52}
\partial_x[\ u_t+6uu_x+u_{xxx}\ ]+\epsilon\ u_{yy}=0 \end{equation}
on application of a class of involutory transformations analogous to $\mathcal{R^*}$ but with $u^*=\rho^{-1}(x)u$ determines an associated system of S-integrable spatially modulated Kadomtsev-Petviashvili equations with
\begin{equation} \begin{array}{l} \label{53}
\partial[\ \rho^{*-1}u^*_{t^*}+6\rho^*u^*\partial(\rho^{*-1}u^*)/\partial x^* \\[3mm]
\qquad\qquad +(\rho^{*2}\partial/\partial x^*)(\rho^{*2}\partial/\partial x^*)(\rho^{*2}\partial/\partial x^*)\ ]/\partial x^* \\[3mm]
\qquad\qquad + \epsilon\ \rho^{*-3}u^*_{y^*y^*}=0\ . \end{array} \end{equation}

Modulated versions of 2+1-dimensional nonlinear evolution equations reciprocally related to canonical S-integrable equations may be generated by conjugation with classes of involutory transformations as embodied in $\mathcal{R^*}$. Thus, in particular, it was established in \cite{cr87} that the S-integrable 2+1-dimensional Dym equation is linked via a reciprocal transformation to the 2+1-dimensional Krichever-Novikov equation
\begin{equation} \label{54}
\frac{\partial}{\partial y}(\phi_y/\phi_x)+\frac{\partial}{\partial x}[\ \phi_t/\phi_x+ \{\phi, x \}+\frac{1}{2}(\phi_y/\phi_x)^{2}\ ]=0 \end{equation}
wherein
\begin{equation} \label{55}
\{\phi, x \}=\frac{\partial}{\partial x}\left(\frac{\phi_{xx}}{\phi_x}\right)-\frac{1}{2}\left(\frac{\phi_{xx}}{\phi_x}\right)^2 \end{equation}
denotes the Schwarzian derivative of $\phi$. Conjugation of this reciprocal link with the application of a class of involutory transformations connects the 2+1-dimensional Krichever-Novikov equation. It is remarked that invariance of the 2+1-dimensional Dym equation under a novel reciprocal transformation has been applied in \cite{crmspv94} to solve classes of initial / boundary value problems for the 2+1-dimensional Krichever-Novikov equation. \\

\subsubsection{Temporal Modulation}

The 2+1-dimensional extended Dym equation
\begin{equation} \label{56}
u_t+2\partial_x(1-\partial_{xx})\left(\frac{1}{u^{1/2}}\right)+6u^2[\ u^{-1}\partial^{-1}_x(u^{1/2})_y\ ]_y=0, \end{equation}
under the action of the class of involutory transformations
\begin{equation} \begin{array}{c} \tag{$\mathcal I^*$}\label{57}
\begin{array}{c} dt^*=\rho^{-2}(t)dt\ , \quad dx^*=dx\ , \quad dy^*=dy \\[3mm]
u^*=u/\rho(t)\ , \quad \rho^*=1/\rho(t) \end{array} 
\end{array} \end{equation}
produces an associated class of S-integrable equations with temporal modulation, namely
\begin{equation} \begin{array}{l} \label{58}
\partial/\partial t^*(\ \rho^{*-1}u^*)+2\rho^{*-3/2}\partial_{x^*}(1-\partial_{x^*x^*})\left(\dfrac{1}{u^{*1/2}}\right) \\[4mm]
\qquad\qquad + 6\rho^{*-5/2}u^{*2}[\ u^{*-1}\partial^{-1}_{x^*}(u^{*1/2})_{y^*}\ ]_{y^*}=0\ . \end{array} \end{equation}

In the case of the Kadomtsev-Petviashvili equation \eqref{53}, application of the involutory transformations $\mathcal{I}^*$ results in an associated S-integrable class with temporal modulation, namely
\begin{equation} \begin{array}{c} \label{59}
\partial/\partial x^*[\ \rho^{*2}\partial/\partial t^*(\ u^*/\rho^*\ )+6(\ u^*/\rho^*\ )(\ u^*/\rho^*\ )_{x^*}+(\ u^*/\rho^*\ )_{x^*x^*x^*}\ ] \\[4mm]
+\ ( \epsilon/\rho^* )u^*_{y^*y^*}=0. \end{array} \end{equation}
It is remarked that, in \cite{cr19} involutory and reciprocal-type transformations have been applied in conjunction to reduce a wide class of nonlinear moving boundary problems incorporating heterogeneity to analytically tractable classical Stefan-type problems. Therein, application was made to the analysis of the evolution of seepage fronts in soil mechanics. \\

 \subsection{Ermakov Modulation} 

Temporal  modulation is set down here of the present 2+1-dimensional extended Dym equation with $\rho^*(t^*)$ in the class of involutory transformations $\mathcal{I}^*$ governed by the classical Ermakov equation
\begin{equation} \label{60}
\rho^*_{t^*t^*}+w(t^*)\rho^*=\mathcal{E}/\rho^{*3}\ , \quad \mathcal{E}\ \epsilon\ \mathbb{R} \end{equation}
with its admitted nonlinear superposition principle
\begin{equation} \label{61}
\rho^*=(c_1\Omega^2_1+2c_2\Omega_1\Omega_2+c_3\Omega^2_2)^{1/2} \end{equation}
where $\Omega_1, \Omega_2$ constitute a pair of linearly independent solutions of the auxilliary equation
\begin{equation} \label{62}
\Omega_{t^*t^*}+w(t^*)\Omega=0\ . \end{equation}
Here, the constants $c_i$, $i=1,2,3$ are such that
\begin{equation} \label{63}
c_1c_3-c^2_2=\mathcal{E}/\mathcal{W}^2 \end{equation}
with $\mathcal{W}=\Omega_1\Omega_{2t^*}-\Omega_{1t^*}\Omega_2$ the constant Wronskian of $\Omega_1, \Omega_2$.

The nonlinear equation \eqref{50} has established physical applications, notably, in the analysis of initial-boundary value problems descriptive of the large amplitude radial oscillations of thin shells composed of Mooney-Rivlin hyperelastic materials and subject to various boundary loadings \cite{crwa89}. The nonlinear superposition principle \eqref{61} may be retrieved via a Lie group procedure as in \cite{crur89}. Therein, application was made to solve an initial value problem for moving shoreline evolution in shallow water hydrodynamics. Lie theoretical generalisation and discretisation of Ermakov-type equations which preserve admittance of nonlinear superposition principles were subsequently derived in \cite{crwspw97}.

Here, with $\rho^*(t^*)$ in the class of involutory transformations $\mathcal{I}^*$ determined by the classical Ermakov equation \eqref{60}, the corresponding class of 2+1-dimensional extended Dym equations with temporal modulation is given by, \textit{in extenso}
\begin{equation} \begin{array}{l} \label{64}
\partial/\partial t^*[\ (c_1\Omega^2_1+2c_2\Omega_1\Omega_2+c_3\Omega^2_2)^{1/2}u^*\ ] \\[2mm]
\qquad +\ 2(c_1\Omega^2_1+2c_2\Omega_1\Omega_2+c_3\Omega^2_2)^{3/4}\partial/\partial x^*(1-\partial_{x^*x^*})\left(\dfrac{1}{u^{*1/2}}\right) \\[4mm]\qquad\quad +\ 6(c_1\Omega^2_1+2c_2\Omega_1\Omega_2+c_3\Omega^2_2)^{5/4}u^{*2}[\ u^{*-1}\partial_{x^*}(u^{*1/2})_{y^*}]_{y^*}=0. \end{array} \end{equation}

\subsection{Ermakov-Painlev\'e Modulation} 

Hybrid Ermakov-Painlev\'e systems were originally derived in \cite{cr14} via symmetry reduction of a multi-dimensional nonlinear Schr\"odinger system which incorporates a de Broglie-Bohm potential. Application was made therein via Ermakov Painlev\'e II symmetry reduction to the analysis of certain transverse wave motions in generalised Mooney-Rivlin hyperelastic materials. Ermakov-Painlev\'e II structure and integrable reduction has subsequently had physical application in such areas as cold plasma physics \cite{crpc18}, Korteweg capillarity theory \cite{crpc17} and the analysis of Nernst-Planck electrolytic system boundary value problems \cite{pacr15}.

Ermakov-Painlev\'e II-IV integrable modulation of coupled 1+1-dimensional solitonic systems of sine-Gordon, Demoulin and Manakov-type was subsequently detailed in \cite{crwsbm20}. Integrable Ermakov-Painlev\'e modulation of established 2+1-dimensional solitonic systems can likewise be derived.

\subsection*{Acknowledgements}

We thank the anonymous reviewer for the careful reading of the manuscript and his/her fruitful suggestions. 

\label{lastpage}

\begin{thebibliography}{99}


\bibitem{rcdh83} Camassa R and Holm D,
An integrable shallow water equation with peaked solitons,
{\em Phys. Rev. Lett.} {\bf 71}, 1661--1664 (1983).

\bibitem{wscr99} Schief W K and  Rogers C,
Binormal motion of curves of constant curvature and torsion. Generation of soliton surfaces,
{\em Proc. Roy. Soc. London A} {\bf 455}, 3163--3188 (1999).

\bibitem{crjk74} Rogers C and  Kingston J G,
Non-dissipative magnetohydrodynamic flows with magnetic and velocity lines orthogonal geodesics on a normal congruence,
{\em SIAM J. Applied Mathematics} {\bf 26}, 183--195 (1974).

\bibitem{jckt88}  Cavalcante J A and Tenenblat K,
Conservation laws for nonlinear evolution equations,
{\em J. Math. Phys.} {\bf 29}, 1044--1049 (1988).

\bibitem{sckt86} Chern S S and Tenenblat K,
Pseudospherical surfaces and evolution equations,
{\em Stud. Appl. Math.} {\bf 74}, 55--83 (1986).

\bibitem{fc93} Calogero F,
Universal integrable nonlinear PDE in Applications of Analytic and Geometric Methods in Nonlinear Differential Equations, Editor: Peter A Clarkson, {\em NATO ASI Series: Mathematics and Physical Sciences} {\bf 413}, 109-114 (1993).

\bibitem{af80}  Fokas A,
A symmetry approach to exactly solvable evolution equations,
{\em J. Math. Phys.} {\bf 21}, 1318--1325 (1980).

\bibitem{cr17} Rogers C,
Moving boundary problems for an extended Dym equation. Reciprocal connection,
{\em Meccanica} {\bf 52}, 3511--3540 (2017).

\bibitem{cr15} Rogers C,
Moving boundary problems for the Harry Dym equation and its reciprocal associates,
{\em Zeit angew. Math. Phys.} {\bf 66}, 3025--3220 (2015).

\bibitem{crpw84} Rogers C and Wong P,
On reciprocal B\"acklund transformations of inverse scattering schemes,
{\em Physica Scripta} {\bf 30}, 10--14 (1984).

\bibitem{crmn86} Rogers C and Nucci M C,
B\"acklund transformations and the Korteweg-de Vries hierarchy,
{\em Physica Scripta} {\bf 33}, 289--292 (1986).

\bibitem{bkvd84} Konopelchenko B G and V.G. Dubrovsky,
Some new integrable nonlinear evolution equations in 2+1-dimensions,
{\em Phys. Lett. A} {\bf 102}, 15--17 (1984).

\bibitem{cr87} Rogers C,
The Harry Dym equation in 2+1-dimensions: a reciprocal link with the Kadomtsev Patviashvili equation,
{\em Phys. Lett. A} {\bf 120}, 15--18 (1987).

\bibitem{cr86} Rogers C,
Reciprocal transformations in 2+1-dimensions,
{\em J. Phys. A: Math. Gen.} {\bf 49}, L491--L496 (1986).

\bibitem{bkvp70} Kadomtsev B B  and  Petviashvili V I,
On the stability of solitary waves in weakly dispersive media,
{\em Sov. Phys. Pokl.} {\bf 15}, 539--541 (1970).

\bibitem{hdmp98} Dai H H and Pavlov M,
Transformations for the Camassa-Holm equation, its high frequency limit and the sinh-Gordon equation,
{\em J. Phys. Soc. Japan} {\bf 67}, 3635--3657 (1998).

\bibitem{vzas74}  Zakharov V E and  Shabat A S,
A scheme for integating the nonlinear equations of mathematical physics by the method of the inverse problem,
{\em J. Funct. Anal. Appl.} {\bf 8}, 226--235 (1974).

\bibitem{bk92} Konopelchenko B G,
{\em Introduction to Multidimensional Integrable Equations. The Inverse Spectral Transform in 2+1-Dimensions}, Technical Editor Rogers C, Plenum Press, New York and London (1992).

\bibitem{pnvz20} Nabelek P V  and  Zakharov V E,
Solutions to the Kaup-Broer system and its 2+1-dimensional integrable generalization via the dressing method,
{\em Physica D: Nonlinear Phenomena} {\bf 409}, 132478 (2020).

\bibitem{crop11} Rogers C and  Pashaev O K,
On a 2+1-dimensional Whitham-Broer-Kaup system: a resonant NLS connection,
{\em Stud. Appl. Math.} {\bf 127}, 141--152 (2011).

\bibitem{wocr93} Oevel W and Rogers C,
Gauge transformations and reciprocal links in 2+1-dimensions,
{\em Rev. Math. Phys.} {\bf 5}, 299--330 (1993).

\bibitem{afcr01} Fokas A S  and Rogers C,
Initial/boundary value problems for simultaneous evolution in three dimensions,
{\em Stud. Appl. Math.} {\bf 107}, 391--401 (2001).

\bibitem{af97}  Fokas A S,
A unified transform method for solving linear and certain nonlinear PDEs,
{\em Proc. Roy. Soc. London A} {\bf 453}, 1411--1443 (1997).

\bibitem{jkcr82}  Kingston  J G and Rogers C,
Reciprocal B\"acklund transformations of conservation laws,
{\em Phys. Lett. A} {\bf 92}, 261--264 (1982).

\bibitem{crws02} Rogers C and Schief W K,
{\em B\"{a}cklund and Darboux Transformations. Geometry and Modern Applications in Soliton Theory},
Cambridge Texts in Applied Mathematics, Cambridge University Press (2002).

\bibitem{cr22} Rogers C,
{\em A Nonlinear Progress to Modern Soliton Theory},
Cambridge Scholars Publishing (2022).

\bibitem{crws82} Rogers C and Shadwick W F,
{\em B\"{a}cklund Transformations and Their Applications},
Academic Press, Mathematics in Science and Engineering Series, New York (1982).

\bibitem{crwa89} Rogers C and Ames W F,
{\em Nonlinear Boundary Value Problems in Science and Engineering},
Academic Press, New York (1989).

\bibitem{crtr20} Rogers C and Ruggeri T,
On invariance in 1+1 dimensional isentropic relativistic gasdynamics,
{\em Wave Motion} {\bf 94}, 102527 (7 pp) (2020).

\bibitem{crtrws20} Rogers C, Ruggeri T and Schief W K,
On relativistic gasdynamics: invariance under a class of reciprocal-type transformations and integrable Heisenberg spin connections,
{\em Proc. Roy. Soc. London} A{\bf 476}, 20200487 (2020).

\bibitem{crtr24} Rogers C and Ruggeri T,
The Euler relativistic gasdynamic system. Chaplygin-K\'arm\'an-Tsien laws and substitution principle,
{\em Rend. Lincei. Mat. Appl.} {\bf 35}, 447--457 (2024).

\bibitem{cacrurao90} Athorne C, Rogers C, Ramgulam U and Osbaldestin A,
A linearisation of the Ermakov system,
{\em Phys. Lett. A} {\bf 143}, 207--212 (1990).

\bibitem{crws18} Rogers C and Schief W K,
Ermakov-type systems in nonlinear physics and continuum mechanics, in {\em Nonlinear Systems and Their Remarkable Mathematical Structures},
pp 541--576, Ed: Norbert Euler, CRC Press (2018).

\bibitem{crmspv94} Rogers C, Stallybrass M P and Vassiliou P,
An invariance of the 2+1-dimensional Harry-Dym equation: application to initial/boundary value problems,
{\em Int. J. Nonlinear Mechanics} {\bf 29}, 145--154 (1994).

\bibitem{cr19} Rogers C,
On Stefan-type moving boundary problems with heterogeneity: canonical reduction via conjugation of reciprocal transformations,
{\em Acta Mechanica} {\bf 230}, 839--850 (2019).

\bibitem{crur89} Rogers C and Ramgulam U, 
A nonlinear superposition principle and Lie group invariance: application in rotating shallow water theory,
{\em Int. J. Nonlinear Mechanics} {\bf 24}, 229--236 (1989).

\bibitem{crwspw97} Rogers C, Schief W K and  Winternitz P,
Lie theoretical generalisations and discretisation of the Pinney equation,
{\em J. Math. Anal. Appl.} {\bf 216}, 246--264 (1997).

\bibitem{cr14} Rogers C,
A novel Ermakov-Painlev\'e II system: N+1-dimensional coupled NLS and elastodynamic reductions,
{\em Stud. Appl. Math.} {\bf 133}, 214--231 (2014).

\bibitem{crpc18} Rogers C and  Clarkson P A,
Ermakov-Painlev\'e II reduction in cold plasma physics: application of a B\"acklund transformation,
{\em J. Nonlinear Mathematical Physics} {\bf 25}, 247--261 (2018).

\bibitem{crpc17} Rogers C and Clarkson P A,
Ermakov-Painlev\'e II symmetry reduction in a Korteweg capillarity system,
{\em Symmetry, Integrability and Geometry: Methods and Applications} {\bf 13}, 018 (2017).

\bibitem{pacr15}  Amster P and Rogers C,
On a Ermakov-Painlev\'e II reduction in three-ion electrodiffusion. A Dirichlet boundary value problem,
{\em Discrete and Continuous Dynamical Systems} {\bf 35}, 3277--3292 (2015).

\bibitem{crwsbm20} Rogers C, Schief W K and Malomed B,
On modulated coupled systems. Canonical reduction via reciprocal transformations,
{\em Communications in Nonlinear Science and Numerical Simulation} {\bf 83}, 105091 (2020).



\end{thebibliography}
\end{document}